%% file: main.tex
\documentclass[sigconf,natbib=true,anonymous=false]{acmart}
\AtBeginDocument{%
  }

\setcopyright{acmlicensed}
\copyrightyear{2025}
\acmYear{2025}
\acmDOI{XXXXXXX.XXXXXXX}
\acmConference[Under Review]{Under Review.}{MMMM DD-DD,
  YYYY}{Location}
\acmISBN{XXX-X-XXXX-XXXX-X/XX/XX}
\usepackage{xcolor}
\usepackage{tikz}
\usepackage{pgfplots}
\pgfplotsset{compat=newest} 
\usepgfplotslibrary{groupplots}

\usepackage[textsize=\scriptsize]{todonotes}
\usepackage[inline]{enumitem}
\usepackage{multirow}
\usepackage{float}
\usepackage{subcaption}
\usepackage{adjustbox}
\usepackage{siunitx}
\input{acronyms}
\newcommand{\best}[1]{{\mathbf{#1}\dagger}}
\newcommand{\comparable}[1]{{\mathit{#1}\ddagger}}

\begin{document}

\title[Hierarchical Multi-field Representations for Two-Stage E-commerce Retrieval]{Hierarchical Multi-field Representations for\\ Two-Stage E-commerce Retrieval}

\input{authors}

\begin{abstract}
\input{sections/01_abstract}

\end{abstract}




\maketitle

\section{Introduction}
\label{sec:introduction}
\input{sections/02_introduction}

\section{Related Work}
\label{sec:related_work}
\input{sections/03_related_work}

\section{Methodology}
\label{sec:method}
\input{sections/04_method}

\section{Experiments}
\label{sec:experiments}
\input{sections/05_experiments}

\section{Results}
\label{sec:results}
\input{sections/06_results}

\section{Further Analysis}
\label{sec:analysis_explainability}
\input{sections/07_insights}

\section{Conclusion}
\label{sec:conclusion}
\input{sections/08_conclusion}




\bibliographystyle{ACM-Reference-Format}
\bibliography{reference}

\clearpage
\appendix

\section{Block-triangular Attention}
\label{app_sec:block_attn}
\input{sections/appendix/A_block_attn}

\section{Datasets}
\label{app_sec:datasets}
\input{sections/appendix/B_datasets}

\section{Hyperparameters}
\label{app_sec:hyperparameters}
\input{sections/appendix/C_hyperparameters}

\end{document}

%% file: acronyms.tex
\usepackage[acronym, nopostdot]{glossaries-extra}
\glsdisablehyper
\setabbreviationstyle[acronym]{long-short}
\newacronym{mvr}{MVR}{MultiView document Representations}
\newacronym{mvg}{MVG}{Multi-View Geometric Index}
\newacronym{mlm}{MLM}{Masked Language Modeling}
\newacronym{attempt}{ATTEMPT}{AspectcontenT TExt Mutual PredicTion}

\newacronym{charm}{CHARM}{Cascading Hierarchical Attention Retrieval Model}
\makeglossaries

\usepackage{xspace}
\newcommand{\madral}{MADRAL\xspace}
\newcommand{\mural}{MURAL\xspace}
\newcommand{\bibert}{BiBERT\xspace}
\newcommand{\ourmodel}{\gls{charm}~}

%% file: authors.tex
\author{Niklas Freymuth}
\authornote{Work done during internship at Amazon.}
\email{niklas.freymuth@kit.edu}
\orcid{0009-0001-7755-6811}
\affiliation{%
  \institution{Kalrsuhe Institute of Technology}
  \city{Kalrsuhe}
  \country{Germany}
}

\author{Dong Liu}
\email{liuadong@amazon.lu}
\affiliation{%
  \institution{Amazon}
  \country{Luxembourg, Luxembourg}
}

\author{Thomas Ricatte}
\email{tricatte@amazon.lu}
\affiliation{%
  \institution{Amazon}
  \country{Luxembourg, Luxembourg}
}

\author{Saab Mansour}
\email{saabm@amazon.com}
\affiliation{%
  \institution{Amazon}
  \country{Santa Clara, CA, US}
}

\renewcommand{\shortauthors}{Freymuth et al.}

%% file: sections/01_abstract.tex
Dense retrieval methods typically target unstructured text data represented as flat strings. 
However, e-commerce catalogs often include structured information across multiple fields, such as brand, title, and description, which contain important information potential for retrieval systems. 
We present \gls{charm}, a novel framework designed to encode structured product data into hierarchical field-level representations with progressively finer detail. 
Utilizing a novel block-triangular attention mechanism, our method captures the interdependencies between product fields in a specified hierarchy, yielding field-level representations and aggregated vectors suitable for fast and efficient retrieval.
Combining both representations enables a two-stage retrieval pipeline, in which the aggregated vectors support initial candidate selection, while more expressive field-level representations facilitate precise fine-tuning for downstream ranking. 
Experiments on publicly available large-scale e-commerce datasets demonstrate that \gls{charm} matches or outperforms state-of-the-art baselines.
Our analysis highlights the framework’s ability to align different queries with appropriate product fields, enhancing retrieval accuracy and explainability.


%% file: sections/02_introduction.tex
Online shopping has become an ubiquitous part of modern life, making it easier to explore product options and quickly find what we need. 
Product retrieval, i.e., the task of surfacing the right products for the right queries, is the backbone of this process and has been a focus of active research~\citep{muhamed2023web, nicholas2024relevance, sen2024text, akshay2024embedding}.
With increasing product diversity and user requirements, product retrieval has faced complex challenges such as diverse search intents~\citep{luo2024exploring}, addressing keyword mismatches~\citep{lakshman2021embracing, priyanka2019semantic} and scaling approaches to work on product corpora spanning millions of items~\citep{sen2024text}. 
Unlike the extensively explored topic of free-form text retrieval, this work focuses on effectively retrieving items that are represented as e-commerce products consisting of structured data.

\begin{figure}[h!]
    \centering
    \includegraphics[width=0.4\textwidth]{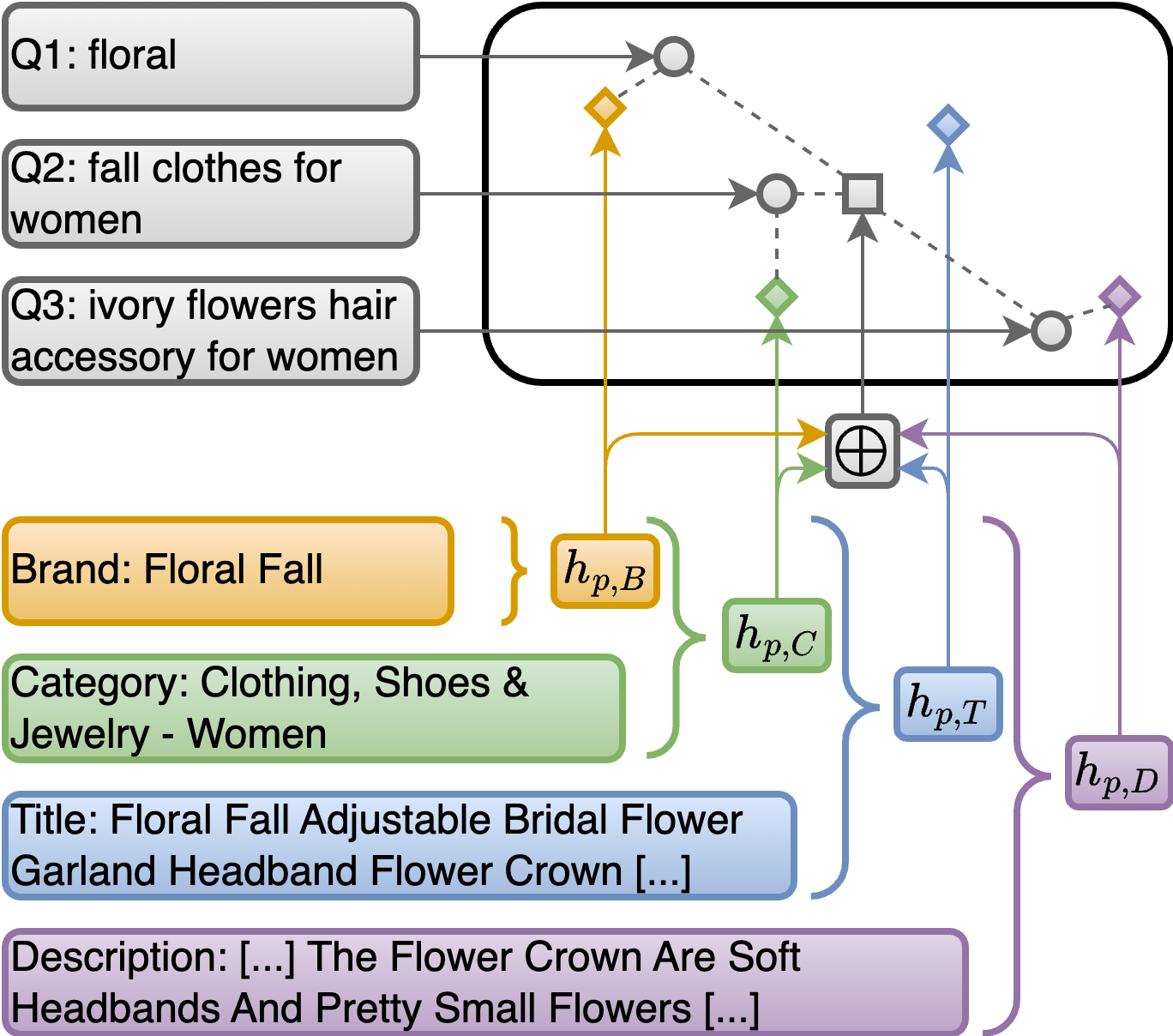}
    \vspace{-0.2cm}
    \caption{
        Retrieval with \gls{charm}.
        Given a product with multiple fields (colored blocks), \gls{charm} produces a cascade of field-level representations (diamonds), with each representation containing information from its own and all previous fields.
        These representations are combined ($\bigoplus$) into an aggregated representation with learned weights, which is used to match query representations (circles) to the product in a first retrieval stage. 
        Queries matched to the aggregated representation are re-evaluated by their distance to the closest field-level representation, allowing for queries with different levels of detail to match the same product.
    }
    \vspace{-0.3cm}
    \label{fig:figure_1}
\end{figure}

On most online stores, a product is defined through multi-faceted product fields, such as its brand, category, title, and descriptions.
Since different customers have different shopping styles and objectives, there are multiple ways to match a product across its various fields, requiring powerful and comprehensive retrieval strategies, as exemplied in Figure~\ref{fig:figure_1}.
While traditional keyword-based approaches like TF-IDF~\citep{salton1988term} and BM25~\citep{robertson2009probabilistic} have been widely used in information retrieval for decades~\citep{baeza1999modern}, the state of the art has changed towards dense information retrieval~\citep{karpukhin2020dense, li2021more, hofstatter2021efficiently, nardini2024efficient}.
However, most approaches in this direction focus on unstructured, plain-text representations, and research specifically addressing multiple product fields largely remains limited to auxiliary reconstruction tasks during model pre-training rather than directly adapting the underlying retrieval~\citep{sun2023attemp, sun2024multi, kong2022multi}. 

In this work, we instead view different product fields as different encodings of the same underlying product with varying degrees of detail.
We propose to consume this hierarchy of product fields to create a cascade of field-level product representations, with each representation accumulating information from the current and all previous fields.
To this end, our method, \glsfirst{charm}, is based on a novel block-triangular attention structure that allows input tokens of one field to attend to all tokens of this field and all previous fields.
\gls{charm} then extracts a representation vector for each product field.
The resulting representations for a product exhibit a naturally emerging diversity across fields, as each representation encodes different levels of detail.
This diversity in turn enables matching the same product to different queries, with `simpler` queries more often matching high-level representations, and more `complex` queries matching more detailed representations.
To reduce retrieval cost, we further combine these field-level representations into an aggregated representation that we use for initial retrieval.
We use this aggregated representation to match a shortlist of promising products per query, and only perform the otherwise costly similarity search between field-level representations and query on this shortlist.
Figure~\ref{fig:figure_1} provides an example of \gls{charm} matching different queries to different fields of the same product.

We experimentally validate our approach on a public collection of large-scale e-commerce datasets~\citep{reddy2022shopping}
.
Here, \gls{charm} outperforms common bi-encoder methods~\citep{Reimers2019SentenceBERTSE, lin2022pretrained}, and matches or outperforms approaches that utilize multiple representations for the same product~\citep{zhang-etal-2022-multi}, while significantly reducing computational cost due to its two-stage retrieval process.
Additional ablation studies show the effectiveness of the individual parts of \gls{charm}.
Finally, we explore how \gls{charm} provides additional explainability through its field-specific matching.
We find increased diversity and complexity for more complex product fields, and strong connections between different kinds of queries and product fields.
We find strong connections between different kinds of queries and product fields, and that more complex product fields yield increasingly diverse representations and query matches.

To summarize our contributions, we
\begin{enumerate*}[label=(\roman*)]
    \item propose a novel block-triangular attention mechanism that allows efficient multi-field processing in e-commerce product retrieval, enabling a cascading hierarchy of field-level product representations.
    \item integrate this mechanism with a two-stage retrieval process to combine fast initial shortlisting with powerful field-level matching.
    \item validate the effectiveness of our approach on several public datasets, matching or outperforming state-of-the-art baselines and providing a detailed analysis of our model's behavior and its inherent explainability.
\end{enumerate*}

%% file: sections/03_related_work.tex
Deep neural networks have significantly advanced the field of information retrieval, with early work using vector representations based on character n-grams processed through multilayer perception models~\citep{huang2013learning}.
In recent years, Transformer models~\citep{Vaswani2017Attention}, and particularly BERT~\citep{devlin-etal-2019-bert}, allowed more effective retrieval based on latent space representations of queries and documents~\citep{karpukhin2020dense, li2021more, hofstatter2021efficiently, nardini2024efficient}.
These methods mostly leverage the vast knowledge and predictive power of pre-trained large language models~\citep{devlin-etal-2019-bert, raffel2020exploring}, enabling a more holistic and semantic retrieval~\citep{hambarde2023information, zhao2024dense}.
In turn, this understanding allows them to significantly improve over classical methods such as TF-IDF~\citep{salton1988term} and BM25~\citep{robertson2009probabilistic} when fine-tuned for information retrieval~\citep{fan2022pre}, as documented in several surveys~\citep{guo2022semantic_review, lin2022pretrained, Li2014semantic}.
These models, such as BiBERT~\citep{Reimers2019SentenceBERTSE, lin2022pretrained}, typically employ contrastive training~\citep{hadsell2006dimensionality, jaiswal2020survey} to structure a latent space that aligns encoded texts by semantic similarity, often within a dual-encoder framework~\citep{bromley1993signature}.
In a typical setting, a large corpus of products or documents is encoded, and evaluation consists of encoding a query and finding the nearest neighbors of this query in the latent space.
Extensions include multitask optimization~\citep{abolghasemi2022improving}, automatic query expansion~\citep{vishwakarma2025fine}, and token-level embeddings that balance expressiveness and computational efficiency~\citep{khattab2020colbert}. 
While not the focus of our work, these extensions could be combined with~\gls{charm} to further improve downstream retrieval performance.
Based on these works, dense retrieval has been successfully applied to e-commerce scenarios~\citep{he2023que2engage, muhamed2023web}, with applications including product search~\citep{magnani2019neural}, click-through rate prediction~\citep{xiao2020deep}, and product ranking~\citep{li2019fromsemantic}.
Yet, most approaches fail to exploit the rich, multi-field structure of data commonly available for e-commerce products.

Recent research has explored multi-field learning to address these challenges. \madral~\citep{kong2022multi} introduces a field-specific module into a dense encoder, generating joint representations for product fields such as color, brand, and category. 
However, this approach relies on categorical labels for each field, which requires pruning of categorical labels and limits its applicability. 
Additionally, the field data is not part of the input of the product encoder, instead influencing its retrieval representation via additional classification tasks.
Building on \madral, \mural~\citep{sun2024multi} aligns field embeddings of different granularities with input token embeddings through self-supervised learning.
Similar to our approach, \mural then combines the resulting embeddings into an aggregated representation using softmax weights that are learned from the product embeddings.
As such, \mural avoids the need for categorical labels, but struggles with complex fields, such as detailed product descriptions, where token embeddings lack semantic depth. 
Other work in this direction~\citep{sun2023attemp} proposes to improve semantic relationships between product fields through mutual prediction objectives. 
Crucially, these methods operate primarily during an additional \gls{mlm} pre-training stage, which addresses the model's lack of attention structure for effective information aggregation~\citep{gao2021condenser}. 
This pre-training phase has been shown to generally enhance the model's capacity to aggregate product information into a single retrieval representation in subsequent contrastive learning~\citep{fan2022pre, gao2021condenser, ma2022pre}, which is further refined through these product-specific auxiliary reconstruction objectives. 
In contrast, our method directly modifies the information processing within the product encoder network by leveraging block-triangular attention.

Another track aims to improve dense retrieval by using multiple vector representations per document or product.
\gls{mvr}~\citep{zhang-etal-2022-multi} trains a single encoder model to produce multiple views of the same input, using a diversity objective between these views to prevent them from collapsing to the same representation vector.
\gls{mvg}~\citep{jiang2022value} applies a similar idea to e-commerce, augmenting the space of latent product representations with clusters of queries that have successfully matched these queries in past interactions. 
In both cases, the retrieval cost increases with the number of document representations. 
While approximate nearest neighbor methods such as Inverted File Index~\citep{sivic2003video} or Hierarchical Navigable Small World Graphs~\citep{malkov2018efficient} can reduce this overhead, \gls{mvg} still requires retrieving a prohibitively large number of items to ensure a fixed number of unique results after de-duplication.

Recent work addresses this challenge by dividing retrieval into two stages~\citep{li2024multi}.
They first generate a shortlist of candidates before refining its order using representations derived from a combination of dense and traditional retrieval methods applied to field-level decompositions of the underlying document~\citep{li2024multi}. 
This two-stage retrieval paradigm is similar to prior work emphasizing retrieve-then-rerank strategies~\citep{guo2022semantic, yates-etal-2021-pretrained, fan2022pre}.
Here, traditional methods often optimize retrieval and reranking independently, limiting their interplay. Joint training approaches \citep{ren2021rocketqav2} bridge this gap by effectively integrating dense retrieval and reranking, but still require training separate retrieval and ranking models.
While we adopt a similar shortlisting strategy, our approach avoids the shallow combination of field-specific representations by constructing hierarchical representations. 
Our approach instead conditions each field-level representation on all preceding fields, enabling richer, more context-aware and naturally diverse embeddings, while requiring only a single encoder forward pass.

%% file: sections/04_method.tex
\subsection{Preliminaries}
Our retrieval pipeline is based on an encoder-only BERT~\citep{devlin-etal-2019-bert}. 
BERT is a transformer-based~\citep{Vaswani2017Attention} model that employs multi-head attention~\citep{bahdanau2015neural}, which allows each token of an input sequence to weigh the importance of other tokens to capture complex contextual relationships. 
For two tokens $i$, $j$, the attention of $j$ towards $i$ is
\begin{equation}
\label{eq:attn}
A_{j}(i) = \text{softmax}\left(\frac{\mathbf{q}_{j} \cdot \mathbf{k}_{i}^T + M_{j, i}}{\sqrt{d}}\right) \cdot \mathbf{v}_{i}\text{,}
\end{equation}
where $\mathbf{q}_{j} \in \mathbb{R}^{d}$ and $\mathbf{k}_{i} \in \mathbb{R}^{d}$ represent the query and key vectors associated with tokens $i$ and $i$, respectively, and $\mathbf{v}_{j} \in \mathbb{R}^{d}$ is the value vector of token $j$. 
The attention mask $M_{i, j}$ is set to $M_{i, j} = 0$ if $i$ is allowed to attend to $j$, and to $M_{i, j} = -\infty$ otherwise. 
By default, BERT utilizes a full attention mask $\mathbf{M}=\mathbf{0}$, allowing each token to attend to all other tokens. 

Given a BERT backbone, we use a dual encoder architecture~\citep{bromley1993signature} to map both product and query encodings to the same latent space using a query and a product encoder with a shared BERT backbone.
We align the vectors of the resulting BiBERT architecture~\citep{Reimers2019SentenceBERTSE, lin2022pretrained} using an InfoNCE loss~\citep{sohn2016improved, oord2018representation}
\begin{equation}
\label{eq:infonce}
    \mathcal{L}_{\text{InfoNCE}}(h_q, h_p) = -\log \frac{\exp(s(h_q, h_{p^+}) / \tau)}{\sum_{i=1}^N \exp(s(h_q, h_{p_i}) / \tau)}\text{,}
\end{equation}
where $h_q$ is the query embedding, $h_{p^+}$ is the positive product embedding and $h_{p_i}$ iterates over all $N$ product embeddings in the batch, including the positive sample $h_{p^+}$, potential hard negative samples~\citep{xiongapproximate}, and in-batch negatives~\citep{karpukhin2020dense}.
We denote the temperature as $\tau$ and use the dot-product for the similarity function~$s(h_q, h_p)$.

Product items typically consist of various fields of information, such as, e.g., a product brand, a title, and a short description. 
These \textit{product fields} contain different levels of information about the product~\citep{reddy2022shopping, zhou2023leveraging}.
The fields naturally form a hierarchy of product representations, where each field can be viewed as a compressed version of the complete product information, with varying degrees of information retention. 
Ordering the fields by their information content, approximated by, e.g., their length, yields an ordered hierarchy of product representations where each new field provides additional and usually more detailed product information than the previous one.
This hierarchy progresses from concise, high-level information, such as color or brand, to more detailed and comprehensive fields, such as the bullet point description, providing a multi-faceted view of the same product that can be leveraged to produce embeddings of varying levels of detail for the same product.

\subsection{Cascading Hierarchical Attention Retrieval Model}
\label{ssec:charm}

\subsubsection*{Block-triangular Attention}

\begin{figure*}[t]
    \centering
    \includegraphics[width=0.75\textwidth]{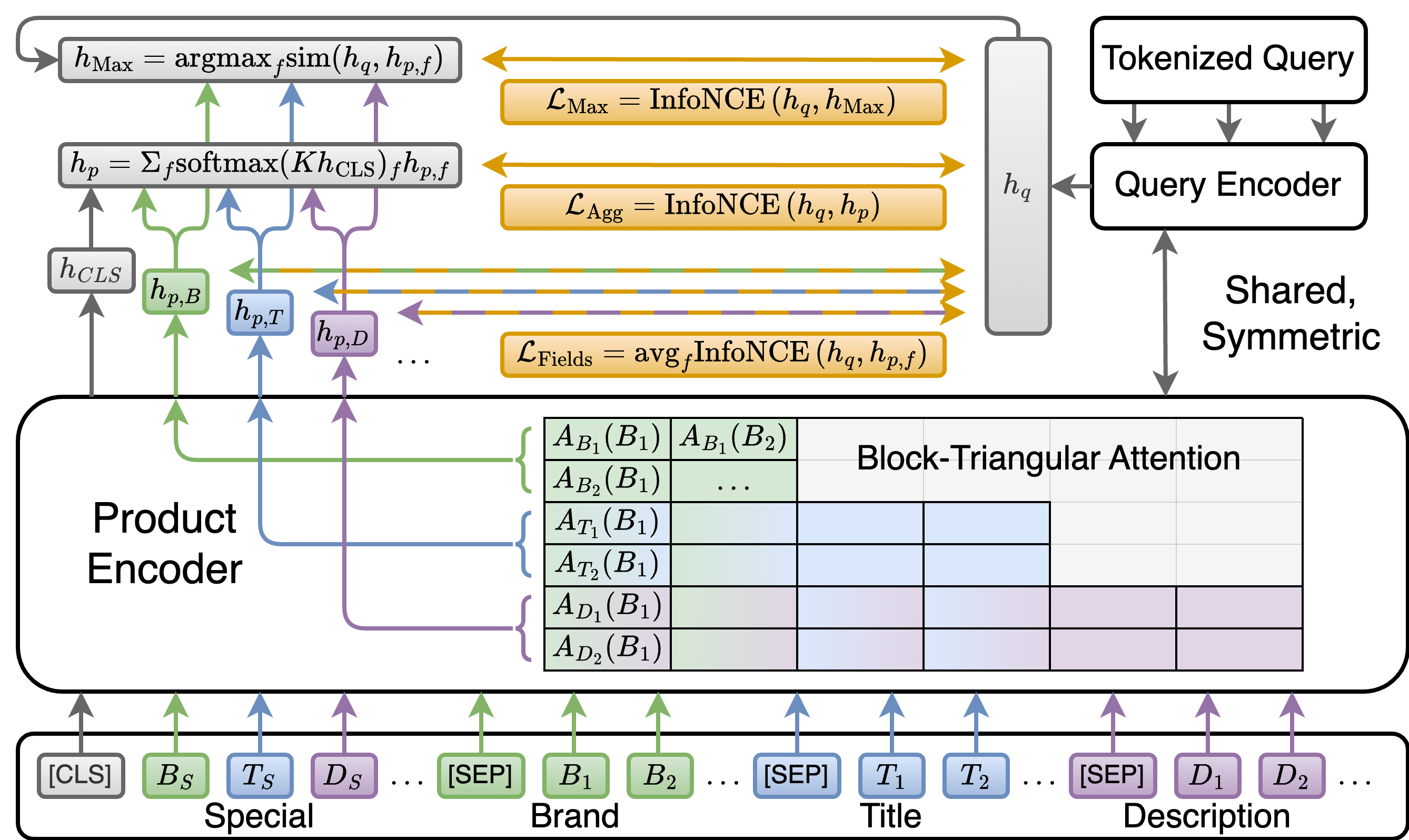}
    \vspace{-0.2cm}
    \caption{
    Schematic overview of \gls{charm}. 
    Our model receives a tokenized product, including a special token per product field. 
    The encoder's attention mask is set to a block-triangular structure that allows tokens to attend to their own field and all previous fields. 
    By arranging fields hierarchically, the field-wise special tokens can capture increasingly detailed product information, enabling diverse product representations in a single model forward pass.
    We train the encoded special tokens and an aggregated representation of them to match the query representation.
    }
    \vspace{-0.2cm}
    \label{fig:schematic}
\end{figure*}

We propose a novel approach to product retrieval that leverages the inherent hierarchical structure of product information. 
Our method, \gls{charm}, generates multiple retrieval vectors for each product, each corresponding to a different level of detail in the product's hierarchical representation.
Unlike previous work that uses diversity losses to differentiate multiple retrieval vectors~\citep{zhang-etal-2022-multi}, \gls{charm} focuses on representing different hierarchies within a product using a single vector for each level. 
Each retrieval vector corresponds to a prefix of product fields, with the first vector representing the top-level field, the second incorporating the next field, and so on. 
This approach allows each vector to capture the residual information provided by its additional field, enabling a dense combination of multiple fields rather than a shallow combination of individual encodings as employed in contemporary work~\citep{li2024multi}.

To implement \gls{charm}, we modify the attention mechanism of our BERT-based encoder. 
Specifically, we modify the attention mask $\mathbf{M}$ to allow attention from each token of one field to other tokens of this field and all previous fields, i.e.,
\begin{equation}
\label{eq:block_attn}
M_{i,j} = 
0 \text{ if } F(i) \geq F(j),\quad 
-\infty \text{ otherwise}
\end{equation}
where $F(i)$ returns the field index for token $i$, and fields are ordered from highest to lowest in the hierarchy.
This \textit{block-triangular attention mask} allows tokens $i$ to attend all tokens $j$ of the current or previous fields, but completely blocks the information flow from all tokens $l$ with $F(l)>F(i)$ to the latent representations of token $i$.
As such, a single forward pass of the model generates a cascade of latent vectors containing increasingly detailed product information.
To capture this information cascade, we add field-wise special tokens to the tokenized input text, and use their final latent representation as the retrieval vector. 
The input $X_p$ thus consists of a special $CLS$ token, followed by $SEP$-separated tokenized fields and their special tokens. 
Appendix~\ref{app_sec:block_attn} provides an example and additional implementation details.

We define the field-level representation vectors as
\begin{equation}
h_{p,f} = \text{BERT}(X_p, \mathbf{M})_f\text{.}
\end{equation}
The subscript $f$ indexes the representation for the special token of field $f\in \mathcal{F}$. 
It also learns the \textit{aggregated representation} $h_p = \sum_f w_f h_{p,f}$, where $w_f = \text{softmax}(Kh_\text{CLS})_f$ with $K:\mathbb{R}^{d\times|\mathcal{F}|}$ is the projected weight for field $f$, similar to previous work~\citep{sun2024multi}.

\subsubsection*{Evaluation}
During evaluation, we first construct a product index by encoding all products using the product encoder. 
This index contains all \textit{field-level representations} $\{h_{p,f}, f \in \mathcal{F}\}$ and the \textit{aggregated representation} $h_p$.
We mirror the special tokens and their aggregated representation for the query tokenization and encoder.
Since we use shared weights between the BERT backbones of both encoders, this symmetry allows our model to more easily provide important information in the representations of the special tokens.

We combine the product index with a two-stage retrieval approach for efficient retrieval.
First, we encode the query and compare it to the aggregated representations $h_p$ of all products, retrieving a shortlist of the $k$ best candidate products.
We then compare the query to all field-wise vectors $h_{p,f}$ of the resulting match set, selecting the highest similarity score over fields for each product.
Doing so only requires a single model forward pass, and can be efficiently implemented using priority queues.
Given $N$ queries and $M$ products, the two-stage approach requires \mbox{$O(N(M + k|\mathcal{F}|))$} comparisons, compared to $O(NM|\mathcal{F}|)$ for full field-dependent retrieval, as done in previous work~\citep{zhang-etal-2022-multi}.
Since usually $M>>k \cdot |\mathcal{F}|$, our two-stage retrieval approach significantly reduces computational cost while maintaining retrieval quality by combining a fast initial retrieval stage with a more expressive second one.
We implement the index using a simple k-Nearest Neighbor search for simplicity, though it can easily be implemented with approximate nearest neighbor methods~\citep{sivic2003video, malkov2018efficient}.

\input{tables/main_results}

\subsubsection*{Training}
\gls{charm} combines multiple InfoNCE losses, as described in Equation~\ref{eq:infonce}, to optimize both the aggregated and field-specific representations.
We match the aggregated representation $h_p$ with the query vector $h_q$ via the loss
$$
\mathcal{L}_{\text{Agg}}=\mathcal{L}_\text{InfoNCE}\left(h_q, h_p\right)\text{,}
$$
ensuring an accurate first retrieval stage.
Additionally, we match the representations of the individual product fields, i.e., 
$$
\mathcal{L}_{\text{Fields}}=\text{avg}_f \mathcal{L}_\text{InfoNCE}\left(h_q, h_{p,f}\right)\text{.}
$$
We finally add an additional loss 
\begin{equation}
    \label{eq:max_loss}
    \mathcal{L}_{\text{Max}}=\mathcal{L}_\text{InfoNCE}\left(h_q, h_{\text{Max}}\right)
\end{equation}
favoring the product field vector $h_{\text{Max}} = \text{argmax}_f \text{sim}(h_q, h_{p,f})$ that most closely matches the query.
Combining these losses, we get
\begin{equation}
\label{eq:full_loss}
\mathcal{L} = \lambda_{\text{Agg}}\mathcal{L}_{\text{Agg}} + \lambda_{\text{Fields}} \mathcal{L}_{\text{Fields}} +  \lambda_{\text{Max}} \mathcal{L}_{\text{Max}}\text{.}
\end{equation}
The last two losses naturally lead to diverse solutions due to the block-triangular attention structure, allowing us to omit explicit diversity losses~\citep{zhang-etal-2022-multi}. 
This structure ensures that the field-level representations have access to different levels of the information hierarchy of the underlying product, resulting in changing ways to match the query as more product information becomes available. 
Each field's retrieval vector is optimized to match the query, with additional emphasis on the best-performing field throughout the optimization process. 
Combined with the loss on the aggregated representation, the total objective encourages the model to learn individually meaningful field-specific representations that can be efficiently combined for a fast first retrieval stage.
Figure~\ref{fig:schematic} provides a schematic overview of the \gls{charm} architecture and its losses.

%% file: tables/main_results.tex
\begin{table*}[t]
\centering
\vspace{-0.1cm}
\caption{
    Performance Comparison of \gls{charm}, \gls{mvr}, MURAL and BiBERT Variants on the Multi-Aspect Amazon Shopping Queries Dataset. 
    Results for MURAL use different pre-training and training hyperparameters and are taken from their paper~\citep{sun2024multi}. Bold$\dagger$ indicates best performance, \textit{italic}$\ddagger$ indicates second best.
}
\vspace{-0.2cm}
\label{tab:main_results}
\resizebox{\textwidth}{!}{%
    \begin{tabular}{l
        m{0.06\textwidth}
        m{0.06\textwidth}
        m{0.07\textwidth}
        m{0.06\textwidth}
        m{0.06\textwidth}
        m{0.06\textwidth}
        m{0.07\textwidth}
        m{0.06\textwidth}
        m{0.06\textwidth}
        m{0.06\textwidth}
        m{0.07\textwidth}
        m{0.06\textwidth}
        }
        &                                     \multicolumn{4}{c}{\textbf{US (English)}}          & \multicolumn{4}{c}{\textbf{ES (Spanish)}}       & \multicolumn{4}{c}{\textbf{JP (Japanese)}} \\
                                                \cmidrule(lr){2-5}                                  \cmidrule(lr){6-9}                                  \cmidrule(lr){10-13}
        Method (Evaluation)                     & R$@10$ & R$@100$ & NDCG$@50$~ & ~P$@10$             & R$@10$ & R$@100$ & NDCG$@50$~ & ~P$@10$             & R$@10$ & R$@100$ & NCDG$@50$~ & ~P$@10$ \\
        \toprule
        \gls{mvr} (Avg.)                        & $31.29$ & $62.89$ & $41.10$ & $48.82$             & $26.19$ & $62.04$ & $41.76$ & $60.00$             & $27.38$ & $57.80$ & $40.93$ & $48.61$ \\
        \gls{mvr} (Best)                        & $33.63$ & $65.92$ & $43.76$ & $50.79$             & $\best{30.45}$ & $\best{67.80}$ & $\best{47.13}$ & $\best{63.45}$             & $\comparable{30.08}$ & $\comparable{61.35}$ & $44.44$ & $50.74$ \\
        \bibert                                 & $33.61$ & $66.38$ & $44.21$ & $50.68$             & $29.18$ & $66.87$ & $46.06$ & $62.65$             & $28.89$ & $60.12$ & $43.17$ & $50.01$ \\
        \bibert-Title                           & $31.70$ & $63.63$ & $42.06$ & $50.00$             & $28.50$ & $64.36$ & $44.59$ & $62.10$             & $29.21$ & $59.68$ & $43.63$ & $50.53$ \\
        \bibert-Simple                          & $28.66$ & $58.76$ & $38.28$ & $47.28$             & $24.93$ & $56.42$ & $39.02$ & $56.54$             & $27.37$ & $55.33$ & $40.52$ & $47.49$ \\
        \mural~\citep{sun2024multi}$^*$         &    & $63.71$ & $42.28$ &                    & & & &                                    & & & & \\
        \mural-CONCAT~\citep{sun2024multi}$^*$  &  & $63.89$ & $42.81$ &             & & & &    & & & & \\
        \hline
        \ourmodel(Agg.)                         & $34.14$ & $\comparable{66.75}$ & $44.69$ & $51.25$             & $29.38$ & $66.53$ & $46.00$ & $62.87$             & $29.37$ & $60.28$ & $43.86$ & $50.37$ \\
        \ourmodel (Best)                        & $\best{34.78}$ & $\best{66.86}$ & $\comparable{45.14}$ & $\best{52.08}$             & $\comparable{30.04}$ & $\comparable{67.26}$ & $\comparable{46.74}$ & $\comparable{63.41}$             & $\best{30.32}$ & $\best{61.58}$ & $\best{45.05}$ & $\best{51.65}$ \\
        \ourmodel (Two-Stage)                   & $\comparable{34.77}$ & $\comparable{66.75}$ & $\best{45.20}$ & $\comparable{51.89}$             & $30.00$ & $66.53$ & $46.54$ & $63.36$             & $30.07$ & $60.28$ & $\comparable{44.53}$ & $\comparable{51.12}$ \\
        \bottomrule
    \end{tabular}
}
\vspace{-0.2cm}
\end{table*}

%% file: sections/05_experiments.tex
\subsection{Datasets}
\label{ssec:datasets}
We experiment on English (US), Spanish (ES) and Japanese (JP) subsets of the public Multi-Aspect Amazon Shopping Queries dataset~\citep{reddy2022shopping}, which provides annotated large-scale real-world e-commerce data.
Each dataset consists of a set of several thousand queries with, on average, $20-29$ associated products per query, as well as an extended corpus of hundreds of thousands of products.
Each query-product pair is classified as an exact, substitute, complementary or irrelevant match.
We follow previous work~\citep{sun2023attemp, sun2024multi} for dense retrieval on this dataset for our training and evaluation setups. 
For each training step, we randomly sample an exact match as the positive example per query, while sampling a product from the remaining labels as a hard negative example. 
The evaluation corpus consists of all products in the respective language.
Table~\ref{tab:dataset_statistics} provides statistics on the number of train and test queries, and the size of the total product corpus.

Each dataset contains multiple product fields, spanning a hierarchy of compressions of the full product.
Ordered by their average length, these are "Color", "Brand", "Title", "Description" and "Bullet points".
We use this order for all models unless mentioned otherwise.
For the US dataset, we use an extended version~\citep{sun2024multi} that annotates each product with an additional "Category" field, which is added between "Brand" and "Title".
We tokenize the queries and products as described in Section~\ref{ssec:charm}, and truncate queries after $64$ tokens and products after $400$ tokens.

\subsection{Implementation Details}
Unless mentioned otherwise, we first perform a simple \gls{mlm} pre-training~\citep{fan2022pre} on the product corpus of the respective dataset to adapt the initial BERT checkpoints to general product data. 
We use the same tokenization and data formatting as in the subsequent contrastive training.
Pre-training details are provided in Appendix~\ref{app_ssec:pretraining_hyperparameters}.

We then initialize the shared BERT backbone for the query and product encoders with the resulting pre-trained checkpoint.
From this checkpoint, we train our model on the loss of Equation~\ref{eq:full_loss}. 
For simplicity, we fix the weights $\lambda_{\text{Agg}} = \lambda_{\text{Fields}} = \lambda_{\text{Max}} = 1$.
Similarly, we omit further tuning, such as annealing the temperature of the InfoNCE loss~\citep{zhang-etal-2022-multi}, to reduce hyperparameter complexity in favor of a more straightforward evaluation of our method.
We train for $200$ epochs with a batch size of $1024$, and use gradient caching for contrastive training~\citep{gao2021scaling} to enable in-batch negatives for this larger batch size.
Appendix~\ref{app_ssec:training_hyperparameters} lists further details on the setup and relevant training hyperparameters.

\subsection{Baselines}
We compare our method to several baselines that use a bi-encoder setup and a BERT backbone.
Here, \glsfirst{mvr}~\citep{zhang-etal-2022-multi} utilizes multiple retrieval representations of the same product for query matching. 
\gls{mvr} employs a regular attention mechanism, treating different representations as channels that process the same underlying product information. 
To prevent representation collapse, \gls{mvr} uses a combined loss function
\begin{equation*}
\mathcal{L}_{\text{MVR}} = \mathcal{L}_{\text{Max}} + 0.01 \mathcal{L}_{\text{Div}}\text{,}
\end{equation*}
where $\mathcal{L}_{\text{Max}}$ is the maximum loss of Equation~\ref{eq:max_loss}, and $\mathcal{L}_{\text{Div}}$ is a local diversity loss
\begin{equation}
\label{eq:div_loss}
\mathcal{L}_{\text{Div}} = -\log \frac{e^{f(q, h{p,\text{Max}})/\tau}}{\sum_f e^{f(q, h_{p,f})/\tau}}\text{.}
\end{equation}
This diversity loss matches the best representation to the query while pushing others away, which is required for~\gls{mvr} to prevent the representations from collapsing. 
In our experiments, we set the number of \gls{mvr} representations equal to the number of product fields, ensuring consistency with our method. 
We evaluate both the best individual representation and the average of all representations, as \gls{mvr} does not inherently use an aggregated representation. 
Notably, \gls{mvr} lacks a two-stage evaluation process, making it impractical to use in large-scale applications with too many representations.

Additionally, we compare to a general BiBERT~\citep{Reimers2019SentenceBERTSE, lin2022pretrained} baseline. 
This baseline uses the $CLS$-token embeddings for both the query and product, and is trained on a simple InfoNCE loss as shown in Equation~\ref{eq:infonce}.
We evaluate three variants of BiBERT, each representing different levels of model complexity and input utilization.
The first variant utilizes all available product fields, providing a "dense" combination of all available information. 
The second variant uses only the product title field (\textit{BiBERT-Title}, which is typically the most informative individual field. 
It shows the performance of the best individual product field, and can be seen as a special case of both \gls{charm} and \gls{mvr} when using only one field.
The final variant represents a naive approach that uses only the product title field and omits the \gls{mlm} pre-training step (\textit{BiBERT-Simple}). 
These BiBERT variants enable us to evaluate the impact of three key factors on retrieval performance, namely (1) the use of multiple product fields versus a single field, (2) the effectiveness of individual field representations, and (3) the contribution of \gls{mlm} pre-training.

We include \mural~\citep{sun2024multi} with their original pre-training and hyperparameters, noting that its auxiliary supervised pre-training objective is orthogonal to our approach and that the results are not directly comparable due to the different training setups. 
Because \mural outperforms \madral~\citep{kong2022multi} in their paper, we do not report results for \madral.

\subsection{Ablation Experiments}
To evaluate which parts of \gls{charm} make it uniquely effective, we conduct additional ablation experiments that change individual aspects of it. 
We look at variants of the attention matrix, employing a full attention matrix, as well as a block-\textit{diagonal} attention variant that changes the inequality of Equation~\ref{eq:block_attn} to an equality. 
The full attention variant shows how our method performs if all representations use the full product information instead of the hierarchy encoded in the block-triangular attention.
The block-\textit{diagonal} attention effectively results in a shallow aggregation over multiple independent product fields, similar to previous work~\citep{li2024multi}, but can not leverage complex connections between different product fields.

We additionally inspect the effect of the different loss terms of Equation~\ref{eq:full_loss}, add the diversity loss of \gls{mvr} to our approach, 
and use the $CLS$ token representation for the query encoder instead of our symmetric setup, where both the query and the product encoder use a softmax aggregation of encoded special tokens.
Finally, we inspect the performance of \gls{charm} when ordering the product fields by their retrieval importance instead of their information content, using Title, Bullet Points, Category, Brand, Description and Color as the field order.

%% file: sections/06_results.tex
\subsection{Metrics}
We evaluate the performance of our two-stage retrieval approach (\textit{Two-Stage}), as explained in Section~\ref{ssec:charm}, using a shortlist of $k=100$ products per query.
Additionally, we report the performance of the best matching field (\textit{Best}) for \gls{charm} and the \gls{mvr}~\citep{kong2022multi} baseline.
For \gls{charm}, we evaluate the aggregated representation (\textit{Agg.}). Is this is not available for \gls{mvr}, we use the average of the field-level representations (\textit{Avg.}) as a substitute.
The BiBERT variants are trained and evaluated on the latent representation of the $CLS$ token (\textit{CLS}).
We calculate Recall$@\{10, 100\}$ (\textit{R$@\{10, 100\}$}) using query-product pairs labelled as "exact" as positive data and all others as negative data.
Following previous work~\citep{reddy2022shopping, sun2024multi}, we also report NDCG$@50$, weighting exact pairs with $1.0$, substitutes with $0.1$, complementary matches with $0.01$, and irrelevant matches with $0.0$.
Finally, we report Precision$@\{5, 10\}$ (\textit{P$@\{5, 10\}$}), which we evaluate by using a classifier model trained to predict if a query-product pair is "exact" or not, allowing us to also take query-product pairs into account that are sensible, but not explicitly labelled as such in the training data.

\subsection{Retrieval Performance}  
\label{ssec:main_results}

Table~\ref{tab:main_results} shows the comparison between \gls{charm}, \gls{mvr}, \mural, and the different BiBERT variants.
We find that \gls{charm} generally performs best across tasks, only being outperformed by \gls{mvr} evaluated on the best representation on the ES dataset.
Further, the performance of the aggregated representation of \gls{charm} is on par with or better than that of a regular BiBERT model, indicating that the block-diagonal attention structure of our model does not hurt its overall expressiveness. 
In contrast, simply taking the average of \gls{mvr} representations yields poor representations, presumably because its diversity loss explicitly pushes non-matching representations away from the match.
The two-stage evaluation of \gls{charm} allows it to efficiently predict a shortlist of promising products, and then rank this shortlist by the closest match to any field representation. 
Since we use $k=100$ products for the shortlist, the Recall$@100$ performance is the same between the aggregated and the two-stage evaluation. 
For the other metrics however, the two-stage retrieval significantly improves performance, outperforming all other evaluations of comparable cost.

\input{tables/ablations}

\subsection{Ablation Results}
\label{ssec:ablations}

Table~\ref{tab:ablations} shows the results for different \gls{charm} ablations evaluated with the two-stage retrieval on the US dataset. 
We find that each loss of Equation~\ref{eq:full_loss} uniquely contributes to the performance of our method, and that there is no gain from adding the diversity loss of Equation~\ref{eq:div_loss}.
In particular, not having a loss for the aggregated representation, i.e., setting $\lambda_{\text{Agg}}=0$, yields a comparatively poor two-stage retrieval shortlist. 
While the top matches, such as R$@10$, are less affected by the initial shortlist, it causes a sharp degradation in performance for R$@100$.
Using a diagonal attention leads to a shallow combination of product fields, similar to previous work~\citep{li2024multi}. 
Since the fields contain different information on different levels of compression, processing them individually and only combining their retrieval representation is insufficient for good performance.
In contrast, using a full attention mechanism allows each representation to use information from all fields, but lacks diversity between the representations, even when adding a diversity loss. 
Using the processed $CLS$ token for the query representation degrades performance, likely due to broken symmetry between the encoders, which makes weight-sharing less efficient.
Finally, ordering the fields by their importance for retrieval slightly decreases performance, presumably because the model benefits from shorter and thus more highly compressed fields early in the hierarchy.

%% file: tables/ablations.tex
\begin{table}[t]
    \centering
    \vspace{-0.1cm}
    \caption{Two-stage evaluation results for \gls{charm} and ablations on the US dataset. We report the performance for \gls{charm} and the absolute difference to it for all ablations.}
    \label{tab:ablations}
    \vspace{-0.2cm}
    \begin{tabular}{lcccc}
        Method & R$@10$ & R$@100$ & NDCG$@50$ & P$@10$ \\
        \toprule
        \textbf{\gls{charm}} & $\mathbf{34.77}$ & $\mathbf{66.75}$ & $\mathbf{45.20}$ & $\textbf{51.89}$ \\
        \hline
        \multicolumn{5}{l}{\textbf{Losses}}\\
        Added $\mathcal{L}_{\text{Div}}$ & $-0.03$ & $+0.02$ & $\sim\hspace{-0.26em}0.00$ & $\sim\hspace{-0.26em}0.00$ \\
        $\lambda_{\text{Max}}=0$ & $-0.13$ & $+0.03$ & $-0.23$ & $-0.12$ \\
        $\lambda_{\text{Fields}}=0$ & $-0.35$ & $-0.52$ & $-0.34$ & $+0.10$ \\
        $\lambda_{\text{Agg}}=0$ & $-1.01$ & $-6.46$ & $-1.83$ & $+0.05$ \\
        \hline
        \multicolumn{5}{l}{\textbf{Attention}}\\
        Diagonal Attention & $-1.36$ & $-1.73$ & $-1.38$ & $+0.67$ \\
        Full Attention & $-0.73$ & $-0.16$ & $-0.75$ & $-1.13$ \\
        (+Added $\mathcal{L}_{\text{Div}}$) & $-0.68$ & $-0.22$ & $-0.74$ & $-1.12$ \\
        \hline
        \multicolumn{5}{l}{\textbf{Misc.}}\\
        Other Field Order & $-0.25$ & $-0.34$ & $-0.34$ & $-0.58$ \\
        Asym. Encoders & $-0.40$ & $-0.16$ & $-0.29$ & $-0.18$ \\
        \bottomrule
    \end{tabular}
    \vspace{-0.2cm}
\end{table}

%% file: sections/07_insights.tex
We investigate the effects of \gls{charm}'s block-triangular attention mechanism for the diversity and matching capabilities of the resulting field-level product representations.
For this analysis, we focus on the evaluation queries and product corpus of the US dataset. 
Unless mentioned otherwise, all evaluations use our two-stage retrieval process, and evaluate the top $10$ products and their associated, most relevant product field for each query.

\subsection{Diversity of Field-level Representations}

\begin{figure}[t!]
    \centering
    \input{figures/evaluation/query_avg_length}
    \vspace{-0.6cm}
    \caption{
    Average length of queries matching a product field by closest dot-product similarity.
    Product fields that are on a higher hierarchy level generally match longer queries.
    }
    \label{fig:analysis_query_length}
    \vspace{-0.1cm}
\end{figure}
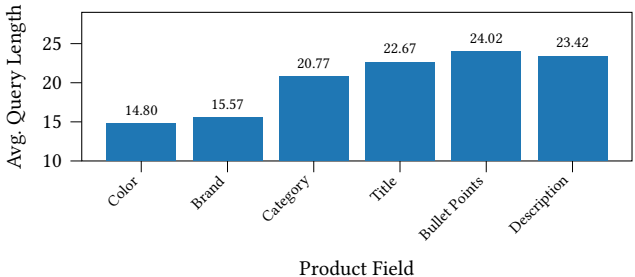

\begin{table}[b] 
    \centering 
    \vspace{-0.1cm}
    \caption{Corpus diversity metrics by product field.} 
    \vspace{-0.2cm}
    \label{tab:product_diversity_metrics} 
    \resizebox{0.47\textwidth}{!}{%
        \begin{tabular}{lccccccc} 
        Metric & Agg. & Color & Brand & Category & Title & Bullet P. & Desc. \\
        \toprule 
        $\uparrow$ Euclidean & 2.6188 & 1.1261 & 1.9845 & 2.9058 & 4.0142 & 4.0671 & 4.0537 \\ 
        $\downarrow$ Dot Product & 19.346 & 19.754 & 19.600 & 19.378 & 19.242 & 19.396 & 19.443 \\
        $\uparrow$ Log-det & -5679.8 & -7411.5 & -6146.6 & -5552.4 & -4916.7 & -4905.9 & -4918.4 \\ 
        \bottomrule 
        \end{tabular} 
    }
    \vspace{-0.2cm}
\end{table}

We analyze the average number of characters in a query that match any given field, using this metric as a proxy for query complexity. 
Figure~\ref{fig:analysis_query_length} demonstrates that longer queries tend to align with representations from longer product fields, suggesting that queries with more complex search intents are more compatible with representations from more detailed product fields.
We further analyze the complexity of the field-level product representations of the whole product corpus by looking at several scalar diversity measures.
We consider the average pairwise euclidean distance (\textit{Euclidean}) between representations of a field, the average pairwise dot-product similarity (\textit{Dot Product}), and the log-determinant of the covariance matrix of the stacked representations of each field (\textit{Log-det}).
Table~\ref{tab:product_diversity_metrics} summarizes our results, finding that representations corresponding to more detailed product fields have higher diversity within the product corpus.
Put together, these results imply that our model learns to produce a cascade of increasingly detailed representations, and matches these with increasingly detailed queries.

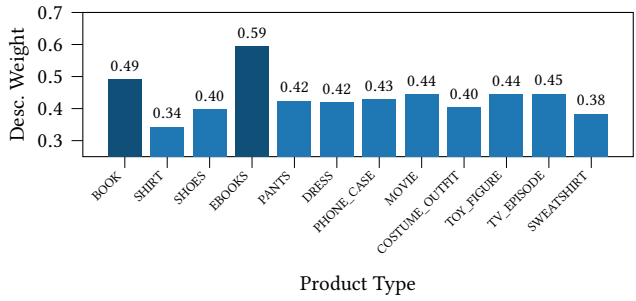
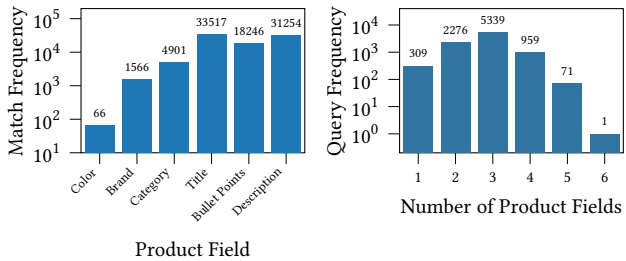
\begin{figure}[t!]
    \centering
    \begin{minipage}[t]{\linewidth}
        \centering
        \input{figures/evaluation/desc_weights}
        \vspace{-0.5cm}
        \subcaption{Weight of the "Description" field in the aggregated representation.}
        \label{fig:desc_weights}
    \end{minipage}

    \vspace{0.2cm} 

    \adjustbox{valign=t}{
        \begin{minipage}[t]{0.48\linewidth}
            \centering
            \input{figures/evaluation/query_field_frequency}
            \vspace{-0.5cm}
            \subcaption{Frequency of product fields appearing as top $10$ matches for any query.}
            \label{fig:query_field_frequency}
        \end{minipage}
    }%
    \hfill
    \adjustbox{valign=t}{
        \begin{minipage}[t]{0.48\linewidth}
            \centering
                \input{figures/evaluation/queries_matching_n_fields}
            \vspace{0.04cm}
            \subcaption{Frequency of queries matching a given number of product fields in their top $10$ matches.}
            \label{fig:queries_matching_n_fields}
        \end{minipage}
    }
    \vspace{-0.1cm}
    \caption{Analysis of field relevance and query matching.}
    \label{fig:combined_analysis}
\end{figure}

Additionally, we explore if the aggregated representation $h_p$ makes effective use of the field-level representations.
For this, we associate product types with all corpus products by crawling the Amazon webstore. 
We then analyze the distribution of softmax weights $w_f$ of the aggregated representation across the most common product categories.
The results in Figure~\ref{fig:combined_analysis}\subref{fig:desc_weights} reveals that media-related products, such as books and e-books, consistently place higher weights on the "Description" field compared to other product types like clothing. 
This aligns with intuitive expectations, as effectively matching media-related products often requires detailed content descriptions.
Put together, these results demonstrates that \gls{charm} captures intuitive and interpretable relationships within a product, and between product fields and queries.
This capability supports the robustness of our approach and lays the groundwork for explainable search systems that dynamically match important product fields.

\subsection{Query-Product Match Analysis}

Figure~\ref{fig:combined_analysis}\subref{fig:query_field_frequency} shows how often each product field is in the top $10$ matches for evaluation queries in the US dataset. 
Match frequency increases with field specificity, with "Title" being the most frequent, likely due to its importance and low noise.
This observation suggests that \gls{charm} often gathers field information up to the "Title" field, while additional fields, such as bullet points or descriptions, may introduce unnecessary information for some queries.

\begin{table}[h]
\centering
\vspace{-0.1cm}
\caption{Qualitative examples of a query matching different products on different fields.}
\vspace{-0.2cm}
\label{tab:query_match_qualitative}
\renewcommand{\arraystretch}{1.5}
\resizebox{0.47\textwidth}{!}{%
    \begin{tabular}{p{0.08\textwidth}p{0.30\textwidth}p{0.20\textwidth}}
    Query & Matched Field & Previous Field \\
    \toprule
    ergonomic desk                          & Category:  Home \& Kitchen - Furniture - Home Office Furniture - Home Office Desks                                                                                                & Brand: EUREKA ERGONOMIC \\
    \cmidrule(lr){2-3}
                                            & Title: RESPAWN RSP-3000 Computer Ergonomic Height Adjustable Gaming Desk [...]                                                                                                    & Category:  Home \& Kitchen - Furniture - Home Office Furniture - Home Office Desks \\
    \cmidrule(lr){2-3}
                                            & Bullet Points:  Go from sitting to standing in one smooth motion with this complete active workstation providing comfortable viewing angles and customized user heights [...]     & Title: VIVO Electric Height Adjustable 43 x 24 inch Stand Up Desk \\ 
    \hline
    pink womans toolbag                     & Category: Tools \& Home Improvement - Power \& Hand Tools - Tool Organizers - Tool Bags                                                                                           & Brand: The Original Pink Box \\
    \cmidrule(lr){2-3}
                                            & Title: Pretty Pink Tool Carry-All With Red Trim-12-1/2 X 9-1/2 X 8 Inches With Multiple Pockets And Metal Handle                                                                  & Category: Tools \& Home Improvement - Power \& Hand Tools - Tool Organizers - Tool Bags \\
    \cmidrule(lr){2-3}
                                            & Bullet Points: Perfect basic set all the essentials are here. Tools and bag are lovely pink with rubbery grips. Great quality tools.                                               & Title: IIT 89808 Ladies Tool Bag 9 Piece \\
        
    \bottomrule                
    \end{tabular}
    \vspace{-0.2cm}
}
\end{table}

Figure~\ref{fig:combined_analysis}\subref{fig:queries_matching_n_fields} examines how many different fields a query matches within the top $10$ results. 
Most queries match $2-3$ fields, suggesting that queries typically span slightly different types of information across products. 
Table~\ref{tab:query_match_qualitative} provides qualitative examples where the same query matches different products through different fields, highlighting the matched field and the previous field in the hierarchy.
For each example, the corresponding matched field provides new information compared to the previous fields, allowing a good match to the query.
Figure~\ref{fig:figure_1} shows the opposite direction, i.e.,  different queries matching the same product on different fields.

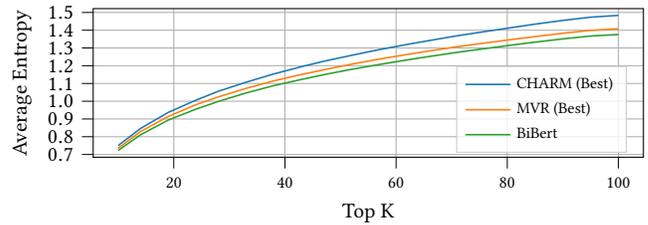
\begin{figure}[t]
    \centering
    \input{figures/evaluation/entropies}
    \vspace{-0.6cm}
    \caption{
    Average entropy of product type distributions across different methods and top-k values
    }
    \label{fig:diversity}
    \vspace{-0.1cm}
\end{figure}

Next, we analyze the diversity of the produced query-product matches at different top $k$ values using the average entropy of product type distributions. 
For each query, we compute the entropy of the product types present in its top $k$ matches, then average these entropies across all queries. 
Higher entropy values indicate greater diversity in the matches. 
Figure \ref{fig:diversity} shows the results for \gls{charm} compared to \gls{mvr} and BiBERT.
\gls{charm} consistently achieves higher entropy values across all examined $k$ values, demonstrating its ability to surface a diverse range of product types.

\subsection{Two-stage retrieval}

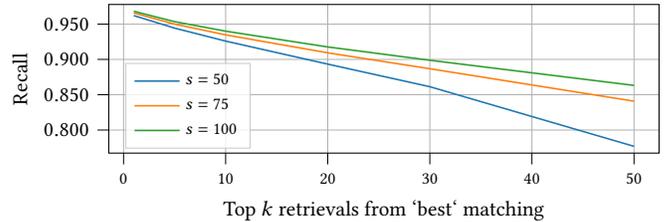
\begin{figure}[t]
    \centering
    \input{figures/evaluation/two_stage_recall}
    \vspace{-0.6cm}
    \caption{Preservation of `best` matches in two-stage retrieval for different initial shortlist sizes $s\in\{50, 75, 100\}$.}
    \label{fig:two-stage-recall}
    \vspace{-0.1cm}
\end{figure}

Figure \ref{fig:two-stage-recall} validates the effectiveness of our two-stage retrieval approach. 
We evaluate how well the first retrieval stage, i.e., using the aggregated representation, preserves the high-quality matches that would be found by directly using the representation of the best matching field. 
Specifically, we analyze what fraction of the ground truth matches from the best matching fields are captured in the initial shortlist. 
We plot recall curves for varying values of top $k$ retrievals, with each curve representing a different shortlist size $s$ from the first stage. 
The results demonstrate strong recall performance. 
For example, with a shortlist size of $50$, over $90\%$ of the `true` top $10$ matches are successfully captured. 
This high preservation of relevant matches validates our two-stage approach, showing we can leverage the efficiency of aggregated representations without significantly compromising retrieval quality.

%% file: figures/evaluation/query_avg_length.tex
\begin{tikzpicture}

\definecolor{darkgray176}{RGB}{176,176,176}
\definecolor{steelblue31119180}{RGB}{31,119,180}

\begin{axis}[
font=\small,
height=0.20\textwidth,
legend style={font=\small},
tick align=outside,
tick pos=left,
width=0.5\textwidth,
x grid style={darkgray176},
xlabel={Product Field},
xmin=-0.69, xmax=5.69,
xtick style={color=black},
xtick={0,1,2,3,4,5},
xticklabel style={font=\scriptsize},
xticklabel style={rotate=45.0,anchor=east},
xticklabels={Color,Brand,Category,Title,Bullet Points,Description},
y grid style={darkgray176},
ylabel={Avg. Query Length},
ymin=10, ymax=29,
ytick style={color=black},
ytick={0,5,10,15,20,25,30},
yticklabels={
  \(\displaystyle {0}\),
  \(\displaystyle {5}\),
  \(\displaystyle {10}\),
  \(\displaystyle {15}\),
  \(\displaystyle {20}\),
  \(\displaystyle {25}\),
  \(\displaystyle {30}\)
}
]
\draw[draw=none,fill=steelblue31119180] (axis cs:-0.4,0) rectangle (axis cs:0.4,14.8030303030303);
\draw[draw=none,fill=steelblue31119180] (axis cs:0.6,0) rectangle (axis cs:1.4,15.5715197956577);
\draw[draw=none,fill=steelblue31119180] (axis cs:1.6,0) rectangle (axis cs:2.4,20.7741277290349);
\draw[draw=none,fill=steelblue31119180] (axis cs:2.6,0) rectangle (axis cs:3.4,22.66709431035);
\draw[draw=none,fill=steelblue31119180] (axis cs:3.6,0) rectangle (axis cs:4.4,24.0169352186781);
\draw[draw=none,fill=steelblue31119180] (axis cs:4.6,0) rectangle (axis cs:5.4,23.4183784475587);
\draw (axis cs:0,14.8030303030303) node[above, font=\scriptsize]{14.80};
\draw (axis cs:1,15.5715197956577) node[above, font=\scriptsize]{15.57};
\draw (axis cs:2,20.7741277290349) node[above, font=\scriptsize]{20.77};
\draw (axis cs:3,22.66709431035) node[above, font=\scriptsize]{22.67};
\draw (axis cs:4,24.0169352186781) node[above, font=\scriptsize]{24.02};
\draw (axis cs:5,23.4183784475587) node[above, font=\scriptsize]{23.42};
\end{axis}

\end{tikzpicture}

%% file: figures/evaluation/desc_weights.tex
\begin{tikzpicture}

\definecolor{darkgray176}{RGB}{176,176,176}
\definecolor{steelblue31119180}{RGB}{31,119,180}
\definecolor{teal1680120}{RGB}{16,80,120}

\begin{axis}[
font=\small,
height=3.5cm,
legend style={font=\small},
tick align=outside,
tick pos=left,
width=1.05\textwidth,
ylabel style={yshift=-0.1cm}, 
x grid style={darkgray176},
xlabel={Product Type},
xmin=-0.99, xmax=11.99,
xtick style={color=black},
xtick={0,1,2,3,4,5,6,7,8,9,10,11},
xticklabel style={font=\tiny},
xticklabel style={rotate=45.0, anchor=east},
xticklabels={
  \(\displaystyle \text{BOOK}\),
  \(\displaystyle \text{SHIRT}\),
  \(\displaystyle \text{SHOES}\),
  \(\displaystyle \text{EBOOKS}\),
  \(\displaystyle \text{PANTS}\),
  \(\displaystyle \text{DRESS}\),
  \(\displaystyle \text{PHONE\_CASE}\),
  \(\displaystyle \text{MOVIE}\),
  \(\displaystyle \text{COSTUME\_OUTFIT}\),
  \(\displaystyle \text{TOY\_FIGURE}\),
  \(\displaystyle \text{TV\_EPISODE}\),
  \(\displaystyle \text{SWEATSHIRT}\)
},
y grid style={darkgray176},
ylabel={Desc. Weight},
ymin=0.25, ymax=0.70,
ytick style={color=black},
ytick={0,0.1,0.2,0.3,0.4,0.5,0.6,0.7},
yticklabels={
  \(\displaystyle {0.0}\),
  \(\displaystyle {0.1}\),
  \(\displaystyle {0.2}\),
  \(\displaystyle {0.3}\),
  \(\displaystyle {0.4}\),
  \(\displaystyle {0.5}\),
  \(\displaystyle {0.6}\),
  \(\displaystyle {0.7}\)
}
]
\draw[draw=none,fill=teal1680120] (axis cs:-0.4,0) rectangle (axis cs:0.4,0.491377047065102);
\draw[draw=none,fill=steelblue31119180] (axis cs:0.6,0) rectangle (axis cs:1.4,0.342143974270231);
\draw[draw=none,fill=steelblue31119180] (axis cs:1.6,0) rectangle (axis cs:2.4,0.398067019618042);
\draw[draw=none,fill=teal1680120] (axis cs:2.6,0) rectangle (axis cs:3.4,0.593994554807429);
\draw[draw=none,fill=steelblue31119180] (axis cs:3.6,0) rectangle (axis cs:4.4,0.424103203166223);
\draw[draw=none,fill=steelblue31119180] (axis cs:4.6,0) rectangle (axis cs:5.4,0.419958396134663);
\draw[draw=none,fill=steelblue31119180] (axis cs:5.6,0) rectangle (axis cs:6.4,0.428738683505926);
\draw[draw=none,fill=steelblue31119180] (axis cs:6.6,0) rectangle (axis cs:7.4,0.444348053103656);
\draw[draw=none,fill=steelblue31119180] (axis cs:7.6,0) rectangle (axis cs:8.4,0.404000495233715);
\draw[draw=none,fill=steelblue31119180] (axis cs:8.6,0) rectangle (axis cs:9.4,0.444012807282861);
\draw[draw=none,fill=steelblue31119180] (axis cs:9.6,0) rectangle (axis cs:10.4,0.446157842406496);
\draw[draw=none,fill=steelblue31119180] (axis cs:10.6,0) rectangle (axis cs:11.4,0.383330130988041);
\draw (axis cs:0,0.491377047065102) node[above, font=\scriptsize]{$0.49$};
\draw (axis cs:1,0.342143974270231) node[above, font=\scriptsize]{$0.34$};
\draw (axis cs:2,0.398067019618042) node[above, font=\scriptsize]{$0.40$};
\draw (axis cs:3,0.593994554807429) node[above, font=\scriptsize]{$0.59$};
\draw (axis cs:4,0.424103203166223) node[above, font=\scriptsize]{$0.42$};
\draw (axis cs:5,0.419958396134663) node[above, font=\scriptsize]{$0.42$};
\draw (axis cs:6,0.428738683505926) node[above, font=\scriptsize]{$0.43$};
\draw (axis cs:7,0.444348053103656) node[above, font=\scriptsize]{$0.44$};
\draw (axis cs:8,0.404000495233715) node[above, font=\scriptsize]{$0.40$};
\draw (axis cs:9,0.444012807282861) node[above, font=\scriptsize]{$0.44$};
\draw (axis cs:10,0.446157842406496) node[above, font=\scriptsize]{$0.45$};
\draw (axis cs:11,0.383330130988041) node[above, font=\scriptsize]{$0.38$};
\end{axis}

\end{tikzpicture}

%% file: figures/evaluation/query_field_frequency.tex
\begin{tikzpicture}

\definecolor{darkgray176}{RGB}{176,176,176}
\definecolor{steelblue31119180}{RGB}{31,119,180}

\begin{axis}[
font=\small,
height=3.5cm,
width=1.12\textwidth,
legend style={font=\small},
log basis y={10},
tick align=outside,
tick pos=left,
x grid style={darkgray176},
xlabel={Product Field},
xmin=-0.5, xmax=5.5,
xtick style={color=black},
xtick={0,1,2,3,4,5},
xticklabel style={font=\tiny},
xticklabel style={rotate=45.0,anchor=east},
xticklabels={Color,Brand,Category,Title,Bullet Points,Description},
y grid style={darkgray176},
ylabel style={yshift=-0.15cm}, 
ylabel={Match Frequency},
ymode=log,
ymin=10, ymax=200000,
ytick style={color=black},
ytick={10,100,1000,10000,100000},
yticklabels={
  \(\displaystyle {10^1}\),
  \(\displaystyle {10^2}\),
  \(\displaystyle {10^3}\),
  \(\displaystyle {10^4}\),
  \(\displaystyle {10^5}\),
}
]
\draw[draw=none,fill=steelblue31119180] (axis cs:-0.4,10) rectangle (axis cs:0.4,66);
\draw[draw=none,fill=steelblue31119180] (axis cs:0.6,10) rectangle (axis cs:1.4,1566);
\draw[draw=none,fill=steelblue31119180] (axis cs:1.6,10) rectangle (axis cs:2.4,4901);
\draw[draw=none,fill=steelblue31119180] (axis cs:2.6,10) rectangle (axis cs:3.4,33517);
\draw[draw=none,fill=steelblue31119180] (axis cs:3.6,10) rectangle (axis cs:4.4,18246);
\draw[draw=none,fill=steelblue31119180] (axis cs:4.6,10) rectangle (axis cs:5.4,31254);
\draw (axis cs:0,66) node[above, font=\tiny]{66};
\draw (axis cs:1,1566) node[above, font=\tiny]{1566};
\draw (axis cs:2,4901) node[above, font=\tiny]{4901};
\draw (axis cs:3,33517) node[above, font=\tiny]{33517};
\draw (axis cs:4,18246) node[above, font=\tiny]{18246};
\draw (axis cs:5,31254) node[above, font=\tiny]{31254};

log basis y={10},
tick align=outside,
tick pos=left,\end{axis}

\end{tikzpicture}

%% file: figures/evaluation/queries_matching_n_fields.tex
\begin{tikzpicture}

\definecolor{darkgray176}{RGB}{176,176,176}
\definecolor{darkslategray66}{RGB}{66,66,66}
\definecolor{steelblue49115161}{RGB}{49,115,161}

\begin{axis}[
font=\small,
height=3.5cm,
width=1.12\textwidth,
legend style={font=\small},
log basis y={10},
tick align=outside,
tick pos=left,
ylabel style={yshift=-0.15cm}, 
unbounded coords=jump,
x grid style={darkgray176},
xlabel={Number of Product Fields},
xmin=-0.5, xmax=5.5,
xtick style={color=black},
xtick={0,1,2,3,4,5},
xticklabel style={font=\scriptsize},
xticklabels={1,2,3,4,5,6},
y grid style={darkgray176},
ylabel={Query Frequency},
ymin=0.2, ymax=40000,
ymode=log,
ytick style={color=black},
ytick={1,10,100,1000,10000},
yticklabels={
  \(\displaystyle {10^0}\),
  \(\displaystyle {10^1}\),
  \(\displaystyle {10^2}\),
  \(\displaystyle {10^3}\),
  \(\displaystyle {10^4}\),
}
]
\draw[draw=none,fill=steelblue49115161] (axis cs:-0.4,0.01) rectangle (axis cs:0.4,309);
\draw[draw=none,fill=steelblue49115161] (axis cs:0.6,0.01) rectangle (axis cs:1.4,2276);
\draw[draw=none,fill=steelblue49115161] (axis cs:1.6,0.01) rectangle (axis cs:2.4,5339);
\draw[draw=none,fill=steelblue49115161] (axis cs:2.6,0.01) rectangle (axis cs:3.4,959);
\draw[draw=none,fill=steelblue49115161] (axis cs:3.6,0.01) rectangle (axis cs:4.4,71);
\draw[draw=none,fill=steelblue49115161] (axis cs:4.6,0.01) rectangle (axis cs:5.4,1);
\addplot [line width=0.9pt, darkslategray66]
table {%
0 nan
0 nan
};
\addplot [line width=0.9pt, darkslategray66]
table {%
1 nan
1 nan
};
\addplot [line width=0.9pt, darkslategray66]
table {%
2 nan
2 nan
};
\addplot [line width=0.9pt, darkslategray66]
table {%
3 nan
3 nan
};
\addplot [line width=0.9pt, darkslategray66]
table {%
4 nan
4 nan
};
\addplot [line width=0.9pt, darkslategray66]
table {%
5 nan
5 nan
};
\draw (axis cs:0,309) node[above, font=\tiny]{309};
\draw (axis cs:1,2276) node[above, font=\tiny]{2276};
\draw (axis cs:2,5339) node[above, font=\tiny]{5339};
\draw (axis cs:3,959) node[above, font=\tiny]{959};
\draw (axis cs:4,71) node[above, font=\tiny]{71};
\draw (axis cs:5,1) node[above, font=\tiny]{1};
\end{axis}

\end{tikzpicture}

%% file: figures/evaluation/entropies.tex
\begin{tikzpicture}

\definecolor{darkgray176}{RGB}{176,176,176}
\definecolor{darkorange25512714}{RGB}{255,127,14}
\definecolor{forestgreen4416044}{RGB}{44,160,44}
\definecolor{lightgray204}{RGB}{204,204,204}
\definecolor{steelblue31119180}{RGB}{31,119,180}

\begin{axis}[
font=\small,
height=0.2\textwidth,
legend cell align={left},
legend style={
  fill opacity=0.8,
  draw opacity=1,
  text opacity=1,
  at={(0.97,0.03)},
  anchor=south east,
  draw=lightgray204
},
legend style={font=\scriptsize},
tick align=outside,
tick pos=left,
width=0.5\textwidth,
x grid style={darkgray176},
xlabel={Top K},
xmajorgrids,
xmin=5.5, xmax=104.5,
xtick style={color=black},
xtick={0,20,40,60,80,100,120},
xticklabel style={font=\scriptsize},
xticklabels={
  \(\displaystyle {0}\),
  \(\displaystyle {20}\),
  \(\displaystyle {40}\),
  \(\displaystyle {60}\),
  \(\displaystyle {80}\),
  \(\displaystyle {100}\),
  \(\displaystyle {120}\)
},
y grid style={darkgray176},
ylabel={Average Entropy},
ymajorgrids,
ymin=0.684602465806917, ymax=1.52168257096505,
ytick style={color=black},
ytick={0.6,0.7,0.8,0.9,1,1.1,1.2,1.3,1.4,1.5,1.6},
yticklabels={
  \(\displaystyle {0.6}\),
  \(\displaystyle {0.7}\),
  \(\displaystyle {0.8}\),
  \(\displaystyle {0.9}\),
  \(\displaystyle {1.0}\),
  \(\displaystyle {1.1}\),
  \(\displaystyle {1.2}\),
  \(\displaystyle {1.3}\),
  \(\displaystyle {1.4}\),
  \(\displaystyle {1.5}\),
  \(\displaystyle {1.6}\)
}
]
\addplot [semithick, steelblue31119180]
table {%
10 0.750402569770813
14 0.846163988113403
19 0.936888217926025
24 1.00615763664246
28 1.05582737922668
33 1.10701942443848
38 1.15369665622711
43 1.19468319416046
47 1.22507619857788
52 1.25861704349518
57 1.29081010818481
62 1.32000553607941
66 1.34193480014801
71 1.36867249011993
76 1.39257884025574
85 1.43419289588928
90 1.45522499084473
95 1.47348523139954
100 1.48363351821899
};
\addlegendentry{\ourmodel (Best)}
\addplot [semithick, darkorange25512714]
table {%
10 0.735047221183777
14 0.828765153884888
19 0.913962960243225
24 0.980669617652893
28 1.02415335178375
33 1.07237768173218
38 1.11490142345428
43 1.15062606334686
47 1.17711293697357
52 1.20870399475098
57 1.23750638961792
62 1.26334488391876
66 1.28330028057098
71 1.30706775188446
76 1.32804310321808
81 1.34848868846893
85 1.36395859718323
90 1.3827338218689
95 1.39921844005585
100 1.40803301334381
};
\addlegendentry{MVR (Best)}
\addplot [semithick, forestgreen4416044]
table {%
10 0.722651481628418
14 0.810385465621948
19 0.893693685531616
24 0.955491304397583
28 0.99845552444458
33 1.04657447338104
38 1.08789813518524
43 1.12281358242035
47 1.14942395687103
52 1.17985200881958
57 1.2068794965744
62 1.23260092735291
66 1.25228571891785
71 1.27538466453552
76 1.2964540719986
81 1.31712472438812
85 1.33279991149902
90 1.35106933116913
95 1.3673369884491
100 1.37570893764496
};
\addlegendentry{BiBert}
\end{axis}

\end{tikzpicture}

%% file: figures/evaluation/two_stage_recall.tex
\begin{tikzpicture}

\definecolor{darkgray176}{RGB}{176,176,176}
\definecolor{darkorange25512714}{RGB}{255,127,14}
\definecolor{forestgreen4416044}{RGB}{44,160,44}
\definecolor{lightgray204}{RGB}{204,204,204}
\definecolor{steelblue31119180}{RGB}{31,119,180}

\begin{axis}[
font=\small,
height=0.20\textwidth,
legend cell align={left},
legend style={fill opacity=0.8, draw opacity=1, text opacity=1, draw=lightgray204},
legend style={font=\scriptsize},
legend pos=south west,
tick align=outside,
tick pos=left,
width=0.5\textwidth,
x grid style={darkgray176},
xlabel={Top $k$ retrievals from `best` matching},
xmin=-1.45, xmax=52.45,
xtick style={color=black},
xtick={-10,0,10,20,30,40,50,60},
xticklabel style={font=\scriptsize},
xticklabels={
  $\displaystyle {\ensuremath{-}10}$,
  $\displaystyle {0}$,
  $\displaystyle {10}$,
  $\displaystyle {20}$,
  $\displaystyle {30}$,
  $\displaystyle {40}$,
  $\displaystyle {50}$,
  $\displaystyle {60}$
},
y grid style={darkgray176},
ylabel={Recall},
ymin=0.767350418760469, ymax=0.977737241764377,
ytick style={color=black},
ytick={0.75,0.8,0.85,0.9,0.95,1},
yticklabels={
  $\displaystyle {0.750}$,
  $\displaystyle {0.800}$,
  $\displaystyle {0.850}$,
  $\displaystyle {0.900}$,
  $\displaystyle {0.950}$,
  $\displaystyle {1.000}$
},
grid=both,
grid style={line width=.1pt, draw=gray!10},
major grid style={line width=.2pt,draw=gray!50},
]
\addplot [semithick, steelblue31119180]
table {%
1 0.962032318115234
5 0.944276928901672
10 0.925907373428345
20 0.893311023712158
30 0.861336231231689
50 0.776913404464722
};
\addlegendentry{$s=50$}
\addplot [semithick, darkorange25512714]
table {%
1 0.965829133987427
5 0.949949741363525
10 0.934718012809753
20 0.90929651260376
30 0.88677453994751
50 0.840828657150269
};
\addlegendentry{$s=75$}
\addplot [semithick, forestgreen4416044]
table {%
1 0.96817421913147
5 0.953433752059937
10 0.93994414806366
20 0.917610287666321
30 0.898771643638611
50 0.863175868988037
};
\addlegendentry{$s=100$}
\end{axis}

\end{tikzpicture}

%% file: sections/08_conclusion.tex
We presented a novel adaptive representation framework for efficiently encoding and retrieving multi-faceted product data in e-commerce. 
Our proposed method, called \glsfirst{charm}, uses a novel block-triangular attention matrix for transformer-based models that allows tokens of one product field to attend to all other tokens of this field and all previous fields.
We combine this structure with an explicit hierarchy of product fields, such as product color, brand, or title, to generate increasingly detailed field-level retrieval representations in a single model forward pass.
We then aggregate these retrieval representations into a single representation, which allows for efficient short-listing of promising query-product matches.
These matches are then re-evaluated by identifying the closest alignment with individual field representations, enabling an efficient retrieval pipeline that delivers high-quality product matches tailored to the diverse requirements of different queries.

The empirical validation of our model highlights the importance of leveraging multiple product fields and the effectiveness of their emerging diversity compared to state-of-the-art baselines.
We validate each component of our model through ablation studies and further show that \gls{charm} fosters diverse, interpretable field representations. 
The model leverages diverse product fields, with deeper fields having more complex representations, and tends to align intricate queries with similarly complex product fields.

In future work, we will investigate extending the block-triangular attention matrix to more general attention graphs, allowing subsets of product fields to attend to arbitrary subsets for more effective and diverse communication between selected fields.
Further, we want to use different retrieval dimensions for the different product fields, matching the amount of retrieval dimensions to the information content of the field to allow for more effective retrieval.

%% file: sections/appendix/A_block_attn.tex
\begin{figure}[h!]
    \centering
    \includegraphics[width=0.37\textwidth]{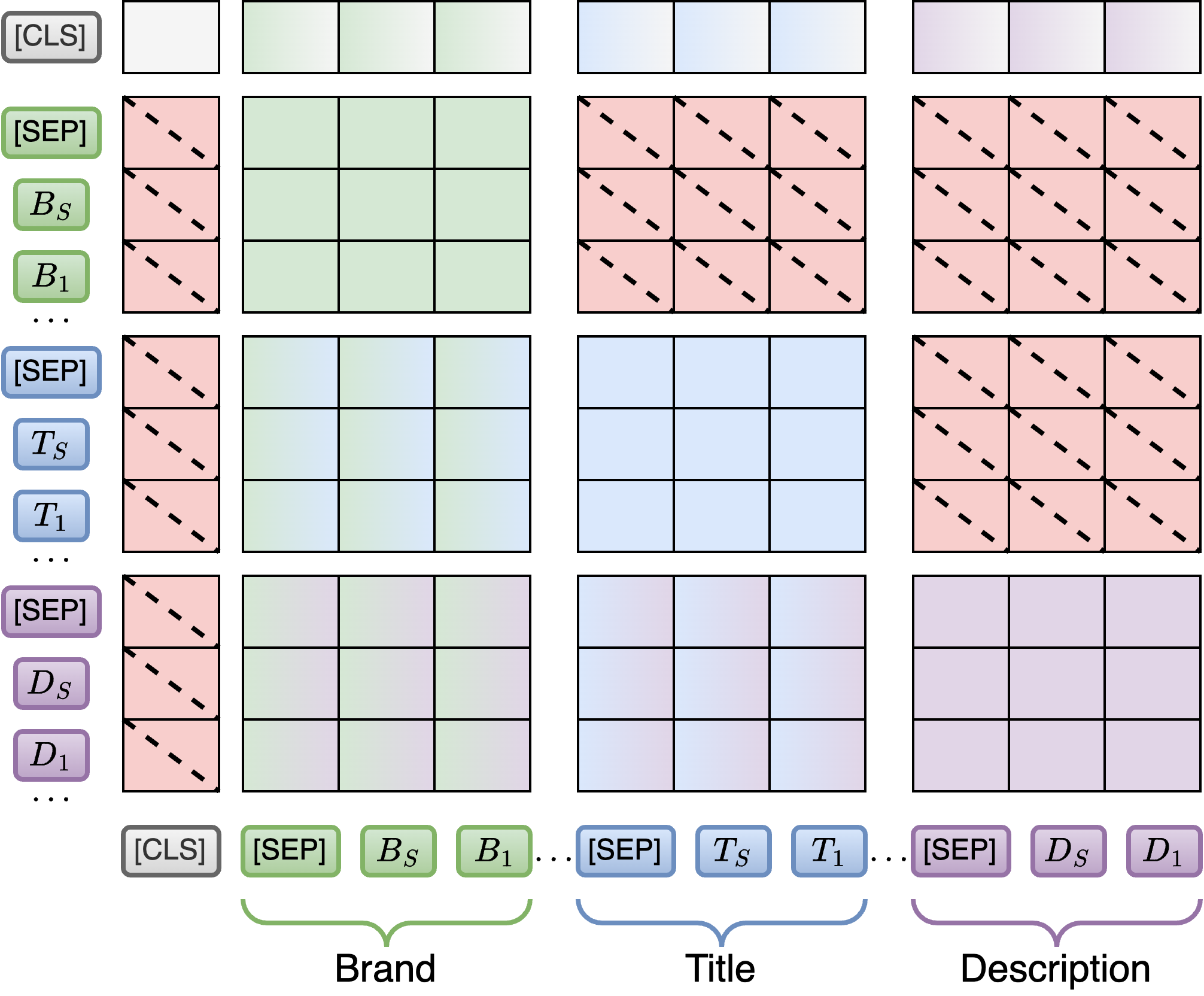}
    \vspace{-0.2cm}
    \caption{
        Exemplary block-diagonal attention matrix.
        Each row ($i$) represents the attention of one token to all tokens in the sequence, while each column ($j$) shows which other tokens a token is attended by.
        The two-colored cells indicate that tokens of one field attend to another field ($M_{i,j}=0$ in Equation~\ref{eq:attn}).
        The red dotted cells indicate masking \mbox{($M_{i,j}=-\infty$)}, which ensures that the tokens of a given field can only attend to tokens of this or previous fields.
        Combined with increasingly detailed fields, this structure yields an information cascade, where the latent vectors of each product field's tokens include increasingly detailed representations.
    }
    \label{fig:block_attn}
    \vspace{-0.2cm}
\end{figure}

Figure~\ref{fig:block_attn} visualizes a block-diagonal attention matrix for exemplary "(B)rand", "(T)itle" and "(D)escription" fields.
In practice, we move all special tokens directly behind the $CLS$ token while maintaining their attention structure to ensure a consistent positional encoding.

%% file: sections/appendix/B_datasets.tex
We provide statistics for the number of train and evaluation queries, their average number of positive and negative product pairs, and size of the full product corpus in Table~\ref{tab:dataset_statistics}. 

\begin{table}[!ht]
    \centering
    \vspace{-0.1cm}
    \caption{Dataset statistics for US, ES, and JP subsets of the Multi-Aspect Amazon Shopping Queries dataset~\citep{reddy2022shopping}}
    \vspace{-0.2cm}
    \label{tab:dataset_statistics}
    \begin{tabular}{lllll}
    Dataset & Type & Amount & Pos. pairs & Neg. pairs \\
    \toprule
    \multirow{3}{*}{US} & Train Queries & 17,388 & 8.70 & 11.41 \\
                        & Test Queries & 8,955 & 8.90 & 11.38 \\
                        & Corpus & 482,105 & -- & -- \\
    \midrule
    \multirow{3}{*}{ES} & Train Queries & 11,336 & 13.44 & 9.77 \\
                        & Test Queries & 3,844 & 12.91 & 11.37 \\
                        & Corpus & 259,973 & -- & -- \\
    \midrule
    \multirow{3}{*}{JP} & Train Queries & 7,284 & 13.20 & 15.51 \\
                        & Test Queries & 3,123 & 13.32 & 15.11 \\
                        & Corpus & 233,850 & -- & -- \\
    \bottomrule
    \end{tabular}
    \vspace{-0.2cm}
\end{table}

%% file: sections/appendix/C_hyperparameters.tex
All model trainings and pre-trainings are conducted using the ADAM~\citep{kingma2015adam} optimizer with a linear learning rate scheduling and a warm-up ratio of $0.1$.
We further train and evaluate using $16$-bit floating point operations, and clip the maximum gradient norm to $1.0$ for all trainings.
Each experiment uses $4$ Nvidia V100 GPUs.

\subsection{MLM Pre-training.}
\label{app_ssec:pretraining_hyperparameters}
Table~\ref{app_tab:pretraining_hyperparameters} provides hyperparameters for the MLM pre-training stage. 
We use the resulting model checkpoints as the initial weights for all experiments unless mentioned otherwise.
We use the same general pre-training parameters across datasets, except that we employ a multilingual BERT (mBERT)~\citep{devlin-etal-2019-bert} model for the non-english ES and JP datasets. 
Since this model is more expensive to run due to an increased token vocabulary, we only train these datasets for $30,000$ steps instead of the $40,000$ for the US one.
\begin{table}[h]
    \centering
    \vspace{-0.1cm}
    \caption{Parameters for the MLM pre-training. Parameters that are only listed once are shared between datasets.}
    \vspace{-0.2cm}
    \label{app_tab:pretraining_hyperparameters}
    \begin{tabular}{lccc}
    & \multicolumn{3}{c}{\textbf{Dataset}} \\
    \cmidrule(lr){2-4}
    \textbf{Parameter} & \textbf{US} & \textbf{JP} & \textbf{ES} \\
    \toprule
    Pretrained checkpoint & BERT (uncased)\footnotemark[2] & \multicolumn{2}{c}{mBERT (cased)\footnotemark[3]} \\
    Training steps & $40{,}000$ & \multicolumn{2}{c}{$30{,}000$} \\
    MLM masking rate & \multicolumn{3}{c}{$\qquad0.15$} \\ 
    Learning rate & \multicolumn{3}{c}{$\qquad1.0 \times 10^{-4}$} \\
    Batch size & \multicolumn{3}{c}{$\qquad512$} \\
    \bottomrule
    \end{tabular}
    \vspace{-0.2cm}
\end{table}

\footnotetext[2]{\url{https://huggingface.co/google-bert/bert-base-uncased}}
\footnotetext[3]{\url{https://huggingface.co/google-bert/bert-base-multilingual-cased}}

\subsection{Training Setup and Hyperparameters.}
\label{app_ssec:training_hyperparameters}
We implement all experiments in pytorch~\citep{paszke2019pytorch}, using the huggingface transformer package~\citep{wolf2020transformers} and Tevatron~\citep{Gao2022TevatronAE} for the contrastive training.
We perform the retrieval using FAISS-GPU~\citep{johnson2019billion, douze2024faiss} with a full similarity search and a dot-product similarity metric.

All contrastive training runs use the final checkpoints from the MLM pre-training stage of the respective dataset as initial model weights.
Table~\ref{app_tab:training_hyperparameters} displays additional training hyperparameters used for \gls{charm} across datasets.
We use the same hyperparameters for all other methods unless mentioned otherwise.
Since the batch size of $1024$ does not fit into memory, we use gradient caching for contrastive training~\citep{gao2021scaling} to allow for all batch samples to act as in-batch negatives for all other samples.

\begin{table}[h]
    \centering
    \vspace{-0.1cm}
    \caption{Parameters for the contrastive training. Parameters that are only listed once are shared between datasets.}
    \vspace{-0.2cm}
    \label{app_tab:training_hyperparameters}
    \begin{tabular}{lccc}
    ~& \multicolumn{3}{c}{\textbf{Dataset}} \\
    \cmidrule(lr){2-4}
    \textbf{Parameter} & \textbf{US} & \textbf{JP} & \textbf{ES} \\
    \toprule
    Training epochs & \multicolumn{3}{c}{$200$} \\
    Learning rate & \multicolumn{3}{c}{$5.0$e$-6$} \\
    Batch size & \multicolumn{3}{c}{$1024$} \\
    $\tau$ (Eq. \ref{eq:infonce}) & $0.1$ & $1.0$ & $1.0$ \\
    $\lambda_{\text{Agg}}$ (Eq. \ref{eq:full_loss}) & \multicolumn{3}{c}{1} \\
    $\lambda_{\text{Fields}}$ (Eq. \ref{eq:full_loss}) & \multicolumn{3}{c}{1} \\
    $\lambda_{\text{Max}}$ (Eq. \ref{eq:full_loss}) & \multicolumn{3}{c}{1} \\
    \bottomrule
    \end{tabular}
    \vspace{-0.2cm}
\end{table}